\documentclass[]{sig-alternate-10pt-mobicom-default}
\usepackage{mathrsfs}
\usepackage{subfigure}
\usepackage{graphics}
\usepackage{balance}
\usepackage{graphicx}
\usepackage{pifont}

\usepackage{amsfonts}

\usepackage{pifont}

\usepackage{epstopdf}

\begin{document}
%

\title{
	Cognitive Radio Networks: Realistic or Not? 
}


\numberofauthors{3}
\author{
\alignauthor Konstantinos Pelechrinis \\
    \affaddr{ School of Information Sciences} \\
		\affaddr{ University of Pittsburgh} \\ 
		\affaddr{ Pittsburgh, PA} \\
		\affaddr{ {\em kpele@pitt.edu} } \\ 
\alignauthor Prashant Krishnamurthy \\
    \affaddr{ School of Information Sciences} \\
		\affaddr{ University of Pittsburgh} \\ 
		\affaddr{ Pittsburgh, PA} \\
		\affaddr{ {\em prashant@sis.pitt.edu} } \\ 
\and
\alignauthor Martin Weiss \\
    \affaddr{ School of Information Sciences} \\
    \affaddr{ University of Pittsburgh} \\ 
		\affaddr{ Pittsburgh, PA} \\
		\affaddr{ {\em mbw@pitt.edu}} \\
\alignauthor Taieb Znati \\
	\affaddr{Department of Computer Science} \\
    \affaddr{University of Pittsburgh} \\ 
		\affaddr{ Pittsburgh, PA} \\
		\affaddr{ {\em znati@cs.pitt.edu} } \\
}

\maketitle
\pagenumbering{arabic}
 
\begin{abstract}
A large volume of research has been conducted in the cognitive radio (CR) area the last decade. However, the deployment of a commercial CR network is yet to emerge. A large portion of the existing literature does not build on real world scenarios, hence, neglecting various important interactions of the research with commercial telecommunication networks.  
For instance, a lot of attention has been paid to {\em spectrum sensing} as the front line functionality that needs to be completed in an efficient and accurate manner to enable an opportunistic CR network architecture. This is necessary to detect the existence of {\em spectrum holes} without which no other procedure 
can be fulfilled.  However, simply sensing (cooperatively or not) the energy received from a primary transmitter cannot enable correct dynamic spectrum access.  
For example, the low strength, or even the absence of detection, of a primary transmitter's signal does not assure that there will be no interference to a nearby primary {\em receiver}. 
In addition, the presence of a primary transmitter's signal does not mean that CR network users cannot access the spectrum since there might not be any primary receiver in the vicinity.  
Despite the existing elegant and clever solutions to the DSA problem no robust, implementable scheme has emerged.  
The set of assumptions these schemes are built upon do not always hold in realistic, wireless environments.  
Specific settings are assumed, which differ significantly from how existing telecommunication networks work.  
In this paper, we challenge the basic premises of the proposed schemes. We further argue that addressing the technical challenges we face in deploying robust CR networks can only be achieved if we radically change the way we design their basic functionalities. In support of our argument, we present a set of real-world scenarios, inspired by realistic settings in commercial telecommunications networks, namely TV and cellular, focusing on spectrum sensing as a basic and critical functionality in the deployment of CRs.  We use these scenarios to show why existing DSA paradigms are not amenable to realistic deployment in complex wireless environments. The proposed study extends beyond cognitive radio networks, and further highlights the often existing gap between research and commercialization, paving the way to new thinking about how to accelerate commercialization and adoption of new networking technologies and services.


\end{abstract}

%
%

\section{Introduction}
\label{sec:intro} 
\setcounter{paragraph}{0}

The increasing demand for wireless connectivity has shifted the attention and efforts of many researchers all over the globe towards the opportunistic dynamic spectrum access paradigm.  
This concept is not new, and was first introduced by Mitola \cite{mitola99}.  
In brief, the idea is that licensed spectrum can be made accessible to unlicensed users (``secondary") when the licensed (``primary") entities are {\em absent}. This absence of a primary user from a specific frequency band at a given point in time and space is referred to as spectrum hole \cite{tandra09}; a reserved portion of the spectrum that is not in use. 
In other words, a spectrum hole is a function of frequency, time and location.  

During the last years there have been significant advancements in hardware technology, enabling engineers to build radios that can {\em understand} their environment and dynamically alter their transmission parameters (e.g., transmission frequency, modulation type etc.). One would have expected that such developments would have lead to large scale cognitive radio network deployments. However, this is not the case, and even a prototype large-scale deployment is yet to appear. In this challenge paper, we argue that this is largely due to the specific mindset we have when we consider the research and design of protocols for CR networks. The majority of the work in this area is theoretical and makes a number of assumptions that may not hold in practice. Even though these studies are arguably quite important and can provide scientific insights, they are not the best avenues to drive practical implementation and eventually commercial adoption and success. 
As we will elaborate in the following sections, specific assumptions that are prevalent in the literature can either expose the primary receiver to harmful interference or limit the performance of a CR network.   

Much of the existing literature focuses on the detection of signals from a single primary transmitter at a location using threshold schemes. If the received signal from this primary transmitter is below a predefined threshold, the band is declared to be vacant. More often than not, a single transmitter/receiver is assumed as the ``secondary" network. As we will see in the following section the vast majority of the studies in this area, are minor variations of the above idea. The main objective of the different schemes is to reduce the sensitivity requirement at the cognitive radio sensor that determines the \textit{absence} of primary transmissions.  

Furthermore, another feature of the existing literature is that it tends to treat the entire frequency spectrum in a unified way. That is, solutions are not different depending on the frequency bands to be used by the CR network.  
However, the nature of the licensed technology in different parts of the spectrum can be significantly different, therefore, requiring very different considerations. For instance, television is a broadcast system with passive receivers, while cellular networks typically include bi-directional asymmetric\footnote{Different frequencies are used for uplink and downlink.} transmissions. Such very different types of primary services create very different kinds of spectrum holes as well, which require different treatment.  

A recent study by Weiss \textit{et al} \cite{weiss10} has classified the resources for a CR network, that is, spectrum holes, based on their characteristics both in space and time. Across time, spectrum holes can be static, periodic or stochastic, while across space they can be contiguous or non-contiguous. The most simple type of spectrum holes that can be used is the temporarily static and spatially contiguous spectrum holes. TV white spaces fall into this category. In particular, after the migration from analog to digital TV, there are unused bands (frequencies  54-698 MHz, which correspond to TV Channels 2-51), and the first coordinated effort to utilize these frequencies for cognitive radio communications has emerged through various proposals (e.g., IEEE 802.22 and the White Spaces Coalition \cite{coalition}). Even though these bands can still be used by unlicensed but authorized users (e.g., microphones and medical telemetry equipment), a temporal static, spatially contiguous spectrum hole can be discovered fairly easily, with a database including all the information for unused analog TV bands across various locations.  In this way, the discovery process takes place offline and does not require the deployment of complex online sensing techniques. It is interesting to see that the one and only effort to commercialize CR networks presented above, i.e., 802.22, does not rely on any spectrum sensing scheme proposed in the research literature. On the contrary, it utilizes a simple, centralized solution. At first glance, it might not appear as the most appealing approach (e.g., from the perspective of its research novelty). However, it is one that can be easily deployed and can successfully drive  deployment.  


Using two representative examples of primary user technologies, that of television and of a cellular network, we argue that current research proposals fail to address ``system" level questions;   
{\bf {\em are the spectrum holes we can identify with existing algorithms really useful?  Why does the presence of a primary signal necessarily render the frequency unusable? How can we identify the regions where passive receptions of a broadcast system exist? How can we take advantage of the different uplink and downlink frequencies in asymmetric systems? }} Similar questions are many times ignored regardless of their importance to the realization of  commercial CR networks. The answers to some of these questions might be simpler than we think (e.g., the {\em database} solution presented above) and/or require research directions that are substantially different from the current literature. We argue through our examples that if we change our mindset and align our thinking more with the way commercial telecommunication networks operate and less in terms of mathematical modeling and protocol design, some of these questions may be answered leading to more rapid development of CR systems.  

The contribution of this paper is twofold.  Firstly, it brings up to the surface, aspects that have been traditionally loosely or inadequately treated in the current literature of cognitive radio networks. However, we argue that these issues are rather important if we envision the deployment of such systems. Unless we start considering them, practical solutions will not be quickly viable. Secondly, and more important, we manifest the disjoint paths that research and industry have followed in the course of the years.  Research should be able to drive commercial deployments.  Nevertheless, this is not the case in the majority of the instances.  The focus of this paper - CR networks - is possibly the most glaring example.  

The rest of the paper is organized as follows. Section \ref{sec:related} briefly discusses representative literature on spectrum sensing.  Section \ref{sec:topology} introduces the two examples we will build our arguments on.  We further elaborate on them in Sections \ref{sec:passive_rx} (broadcast TV system) and \ref{sec:fdd_sys} (cellular network).  Finally, Section \ref{sec:scope} discusses the scope of and concludes our study.  

\section{Related Studies}
\label{sec:related} 
\setcounter{paragraph}{0}

In this section we will briefly describe the approaches that are currently proposed for spectrum sensing\footnote{The list of papers presented in this section is by no means exhaustive due to space limitations.  We are only interested in presenting the general approaches used in spectrum sensing in order to build on them our discussion in the following sections.  The interested user can see \cite{tandra09} \cite{survey-cognets} for additional references.}.  
Despite the fact that efficient spectrum sensing has been identified as key to the success of cognitive radio networks very early \cite{sahai04}, we still lack a satisfactory approach to perform this task. Existing literature can be broadly categorized into three classes: (i) non-cooperative transmitter detection, (ii) co-operative transmitter detection and (iii) interference-based detection.  

{\bf Non-cooperative detection: }
This is the most basic form of spectrum sensing where the secondary transm	itter tries to decide whether there is a primary transmitter using the spectrum or not.  
The detection problem is formulated using hypothesis testing, where the null hypothesis is the absence of the primary transmitter and the alternate hypothesis indicates its presence \cite{digham07}.  
The baseline of this single-user sensing scheme is that of energy detection.     
The secondary transmitter measures the average energy on a specific channel and the decision is based on a threshold comparison (e.g., \cite{li08} \cite{tandra08}).  
However, more accurate detection is possible if the secondary user has information on the primary user's signals.  
In this case, {\em matched filter detection} can maximize the SNR and perform optimally \cite{sahai04}.  
Nevertheless, given that a priori knowledge of the primary signal is not always available, other approaches such as {\em cyclostationary feature detection} (e.g., \cite{cabric04} \cite{fehske05} \cite{haykin08}) and {\em eigenvalue based detection} (e.g., \cite{zeng07} \cite{zeng07a}) are proposed.  
In all of these schemes, every secondary transmitter operates in isolation and takes his decision based on his own measurements.  
Non-cooperative spectrum sensing has been criticized because of its inherent stringent requirement for high reception sensitivity; 
secondary transmitters need to be able to detect primary signals at their circuitry with the received signal strength (RSS) as low as the noise level.  
Hence, a second class of algorithms used to identify spectrum holes includes cooperation among many secondary users as discussed below.  

{\bf Cooperative detection: } 
Non-cooperative spectrum sensing approaches are subject to high uncertainties due to wireless propagation effects. For instance, fading can cause large degradations at the received signal strength from a single radio. Obtaining more measurements from a larger number of co-located radios can provide us with a more robust decision. This is the goal of a cooperative spectrum sensing scheme. There are two basic approaches followed in cooperative detection, namely (a) soft combining and (b) hard combining (e.g., \cite{ganesan05} \cite{sun07} \cite{visotsky05}).  
In soft combining systems, CRs report their actual signal strength measurements to a fusion center, while in the latter case they report their binary decision (as per the non-cooperative approach).  
The fusion algorithm can be either a logic AND, a logic OR or a majority vote based scheme.  
There are different angles from which one can see the benefits of collaborative sensing.  
For instance, collaboration can reduce the sensitivity requirements for individual radios \cite{mishra06}.  
It can also be viewed as means for reducing the time required for sensing or reducing the false alarms rate \cite{ganesan07} \cite{ma07}.  

Regardless of the improvements over non-cooperative schemes, the above systems still suffer from some major limitations.  
There is still lack of information with regards to the location of the primary receiver.  
Even though a primary transmitter's signal can be detected with larger accuracy, interference avoidance is still not guaranteed at the primary receiver.  
Furthermore, spectrum sensors need to be co-located in an fairly small area, in order to assure that the path loss component of the received signal from the primary transmitter is the same for all practical purposes for all sensors (recall that cooperation aims into removing the randomness imposed on a single measurement from wireless effects such as shadow fading).  
If this is not the case, then the performance of the cooperative schemes will be poor.  
If sensors are distributed over a larger area the different path loss components of the signal strength will (legitimately) yield many inconsistencies among the measurements of the single radios.  
This falls back to one of the questions posted in our introduction.  
What is the purpose or {\em value of the spectrum hole} identified from a scheme with these requirements?  
Since primary systems are usually deployed in high-valued areas (e.g., a broadcast TV system needs to serve populated areas), a cooperative sensing system needs to be deployed further.  
Therefore, any spectrum hole identified will have a decreased value (e.g., located in rural areas) imposing also limitations to the applications possible with the underlying CR network.  
We will further elaborate on this in this paper later.  
Finally, even though this paper is not focused on security aspects, we should also consider trust-related issues present in every collaborative scheme; colluding malicious entities can manipulate the outcome of the fusion algorithm. Thus, an additional design constraint, that of robust decision, needs to be satisfied (e.g., \cite{chen08} \cite{parvin11} \cite{kaligineedi10}).

{\bf Interference-based detection: }
Both cooperative and non-cooperative detection do not take into consideration the presence of a primary receiver. However, since interference takes place at the receiver, identifying its presence and/or position is important. The idea of {\em Interference temperature (IT)} has been proposed by the FCC \cite{fcc-d}, to capture the additional interference (above the noise floor) that a primary receiver can tolerate. This means, that using the interference temperature model, we can have simultaneous transmissions from a primary and secondary user, as long as the interference from the latter is beyond the IT of the primary receiver. Even though as a concept this is appealing, there are many practical issues related with this approach. For instance, how is it possible to measure the interference at the primary receiver without prior knowledge of its position relative to the secondary transmitter \cite{brown05}? A promising approach was presented by Wild and Ramchandran \cite{wild05} where low cost sensor nodes are mounted on primary receivers to detect the local oscillator leakage power emitted by the RF front-end of the primary receiver and hence, the presence of the latter. This information is consequently sent as feedback to the secondary users. However, such approaches have received low attention from the community and have been practically abandoned as {\em unrealistic}. The main argument against such proposals is that they require large scale infrastructural upgrades. Nevertheless, as we will argue in Section \ref{sec:passive_rx}, if we want to have a real-world cognitive radio network deployment, this is a promising direction that needs revisiting.  

\section{Topologies for CR Networks}
\label{sec:topology} 
\setcounter{paragraph}{0}

\begin{figure}[htbp] 
   \centering
   \includegraphics[width=\columnwidth]{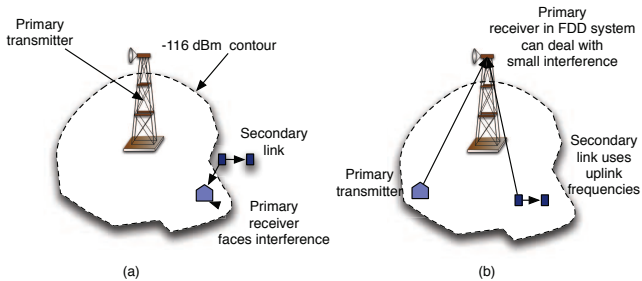} \vspace{-0.2in}
   \caption{Two Topologies and Scenarios for Investigating Challenges with CR Networks}
   \label{fig:fig1}
   \vspace{-0.2in}
\end{figure}

In this paper, we will use two specific topologies and corresponding scenarios to illustrate the problems with existing approaches and need for thinking outside the box for the success of CR networks.  These topologies are shown in Figure~\ref{fig:fig1}.

\subsection{TV White Spaces}
As our first example, we consider the most widely discussed spectrum for usage in a CR network setting which is that of broadcast television white spaces. The corresponding topology and scenario is shown in Figure~\ref{fig:fig1}(a). The IEEE 802.22 standard states that any sensing algorithm used should be able to detect digital television signals at -116 dBm, with probabilities of false alarm and mis-detection, both equal to 0.1. Based on this requirement, if no signal is detected the spectrum is declared as free and secondary transmissions can be scheduled.

This causes two problems that we outline below. First, the signals from the digital TV transmitter can be received with high quality (i.e., RSS $>>$ - 116dBm) at distances of several miles \cite{fcc-data}\footnote{The actual distance is determined by many factors such as the TV tower location and transmission power, the TV receiver/antenna location, etc. but 70 miles is a typical distance.} and  TV signals with very low RSS on the range of -116 dBm will only be received in areas much further from the TV tower. This observation is important, since it practically restricts the cognitive radio network to be in locations that are very far from residential areas. The question that comes to mind is whether there are commercially viable applications that could benefit from a CR network located in an area devoid of human presence. After some thought, one can conclude that there might exist applications that can take benefit from a cognitive radio network architecture in such places (e.g., road safety applications with road side units operating in these bands or sensor networks in forest areas). However, one could also counter-argue that in these areas, unlicensed bands can be as effective as  opportunistic usage of licensed bands as the demand is likely to be low in such locations. 

Except in the case of carrier sensing based wireless networks where a transmitter has to back-off when it senses a competing transmission, interference affects only the reception of the signal in other communication scenarios. This causes the second problem here. As shown in Figure~\ref{fig:fig1}(a), it is possible that the -116 dBm contour is not circular due to radio propagation vagaries. Although a secondary transmitter may, with quite complex processing, detect the signal level from the primary transmitter to be below this value, it has no knowledge of the location or distance of a passive primary receiver (a television). Thus, a secondary transmission can cause significant interference to a receiver. Unless we can detect a primary receiver that is actively receiving the broadcast signal, secondary transmissions may be harmful to the primary system. We investigate this further in Section~\ref{sec:passive_rx}.

\subsection{A Cellular FDD Network}
As the second example, we consider a cellular network that employs frequency division duplexing (i.e., there are separate frequencies for the uplink and downlink). In the US, the uplink (824-849 MHz) and downlink frequencies (869-894 MHz) are separated by 45 MHz in one commonly used slice of licensed spectrum. In a given cell, on the uplink frequencies, only the base station is the receiver as shown in  Figure~\ref{fig:fig1}(b). Channels that are 1.25 MHz (5 MHz) wide are employed in 3G  systems based on the CDMA200  (UMTS) standard. 

It is also well known that the load in cells varies significantly over time and day of week (e.g., \cite{peng11} \cite{Tip10}). Since CDMA systems are interference limited, a secondary transmission that employs \textit{only the uplink frequencies} may be able to operate without causing any harm to the primary system \cite{Peha10}. Especially if the secondary transmissions are employing low transmit power for short range links, there may be negligible interference or harm. This is further helped by the fact that the path loss for short distances usually has a smaller exponent compared to large distances. Thus, while the signal attenuates significantly and has very small power by the time it reaches the primary receiver (the base station), it is sufficient to perhaps support high data rates at shorter distances.

However the cellular spectrum cannot be used for secondary applications as of now. Even if this was the case, if the spectrum is sensed, it will be discovered to be \textit{occupied} by the primary users. The transmitters in this case are the mobile phones which will all employ the same frequency in a CDMA system. Thus, a very small number of mobile phones transmitting on the uplink would still tag the spectrum as occupied making it unusable by a secondary system even when it is unlikely to cause any harm at the primary receiver. The quality of the secondary link will be impacted by the interference from  primary transmissions. We further consider this scenario in Section~\ref{sec:fdd_sys}.

\section{The Passive Receiver Problem}
\label{sec:passive_rx} 
\setcounter{paragraph}{0}

In the previous section we have discussed scenarios where spectrum sensing algorithms as envisaged in the literature {\em fail} to address the real problem; 
either the secondary user will miss opportunities since the interference to the primary receiver is not harmful (FDD network or no receivers exist in the area) or will cause harmful interference since the detection of a primary transmitter does not reveal anything for the location of the primary receiver (TV broadcasts). Even though both of the problems are related to the complex, dynamic and unpredictable topology of commercial wireless networks, broadcast systems (such as TV) exhibit a particular characteristic that will render any current (collaborative or not) sensing algorithm ineffective. In brief, a broadcast system's receivers are {\bf passive}, that is, there are no signs of their presence. Any appropriate receiver that is in range of the broadcast transmitter can receive the signal; if the secondary user is under the coverage of the primary transmitter he will sense the latter's signals. Thus, it appears that a cognitive radio network utilizing the TV spectrum is not possible in urban, residential areas, which have strong TV coverage.

However, as previously mentioned, despite the presence of a primary broadcast signal, if no primary receiver is in the near vicinity (in our case if no TV receiver is turned on), the secondary user can be allowed to occupy the spectrum without harm to the primary system.  Therefore, in broadcast systems like the TV, current sensing algorithms will result in many {\em missed opportunities} for the secondary user. To assess the viability of this claim, we have conducted a large scale user study to identify the {\em usual} times that people have their TV receivers turned on and off and hence, identify periods that opportunistic access is possible. We would like to reiterate here that during the ``off" periods, even though a cognitive sensor would sense broadcast primary signals, secondary transmissions would be possible since the receiver is not powered on and thus not harmed. 

We have conducted two different surveys to identify whether spectrum opportunities exist.  
For the first survey, we have utilized the Amazon's Mechanical Turk platform to recruit people willing to participate in an anonymous survey for a small amount of money. The purpose of this survey is to obtain a general view of the usage patterns of people (e.g., times that the receivers are off, type of TV service, etc.). The second survey is of smaller scale and was conducted with a paper questionnaire distributed to the households of a single apartment building. The purpose of this study is to identify whether similar patterns of TV usage exist in a concentrated residential area. Since in reality there are many receivers within a geographical cluster, we want to study the cumulative activity patterns of co-located users. This will reveal any opportunities for secondary user access in the vicinity of these users.

\begin{figure*}[ht]
\begin{center}
\parbox{2in} {
\centerline{
\includegraphics[scale=.25]{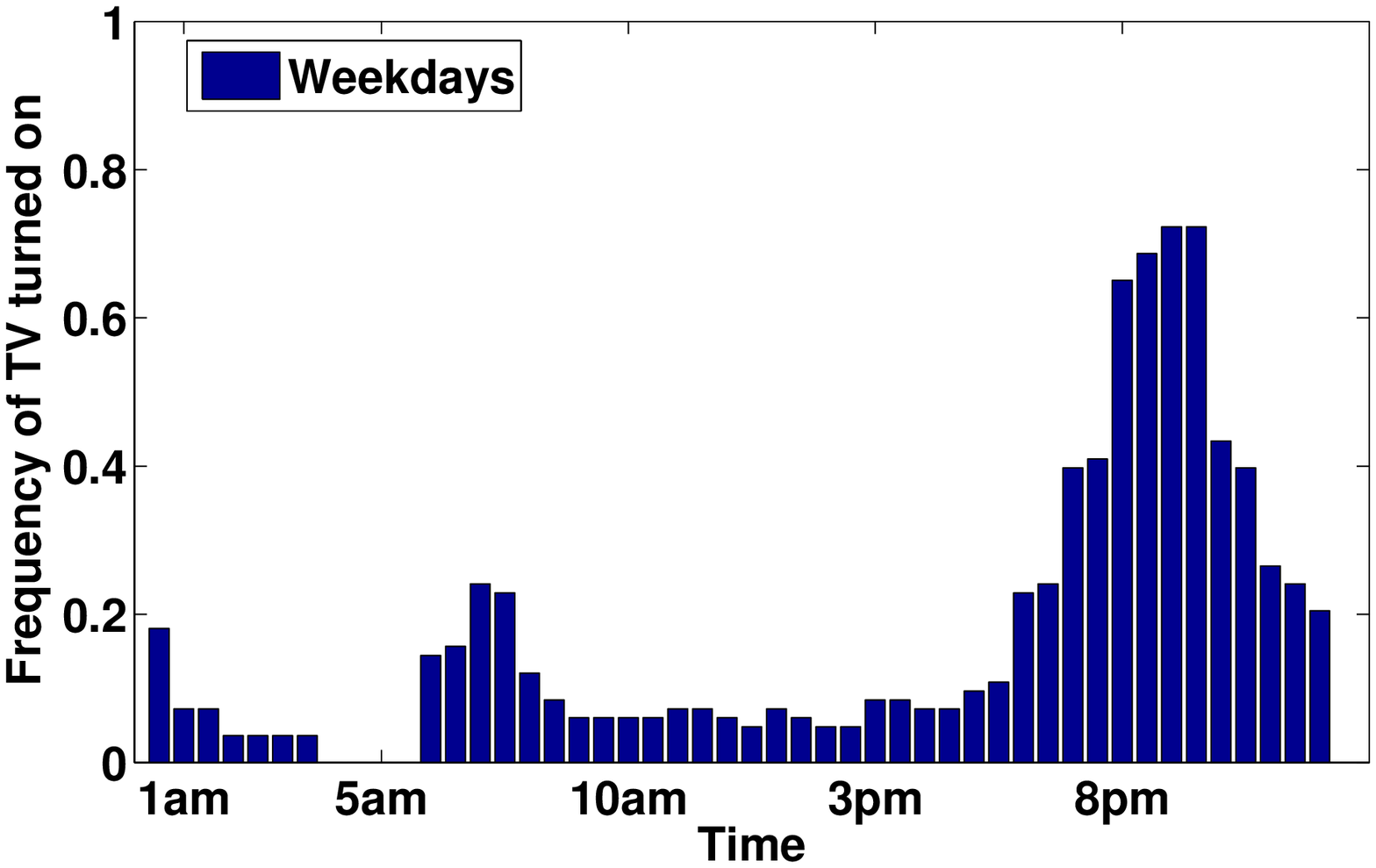}} \vspace{-0.2in}
\caption{TV receiver activity in bins of 30 minutes for a {\em typical weekday}.} \label{fig:t-weekdays}
}
\makebox[.25in] {}
\parbox{2in} {
\centerline{
\includegraphics[scale=.25]{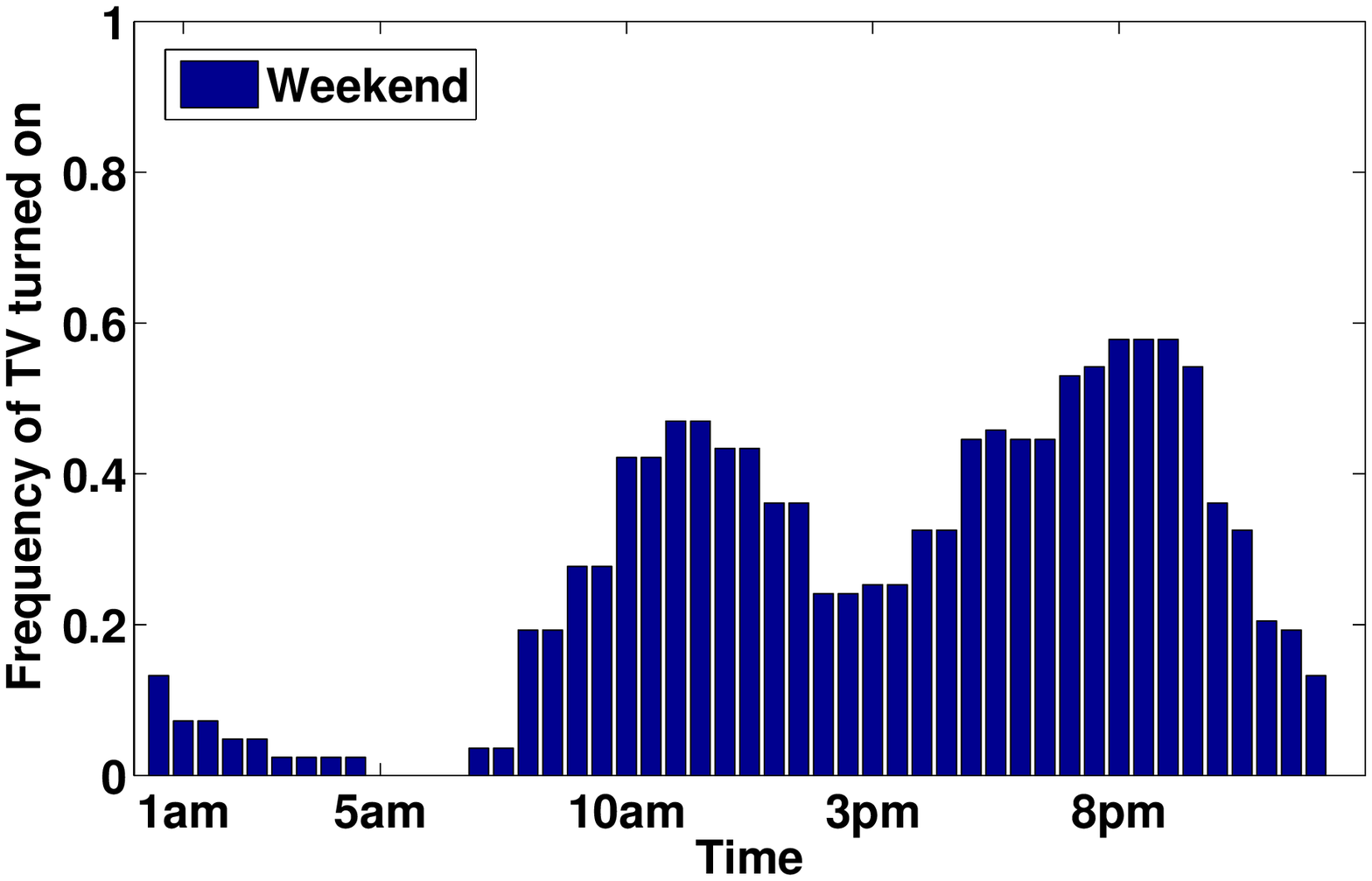}} \vspace{-0.2in}
\caption{TV receiver activity in bins of 30 minutes for a {\em typical weekend}.} \label{fig:t-weekends}
}
\makebox[.25in] {}
\parbox{2in} {
\centerline{\hspace{-0.4cm}
\includegraphics[scale=.25]{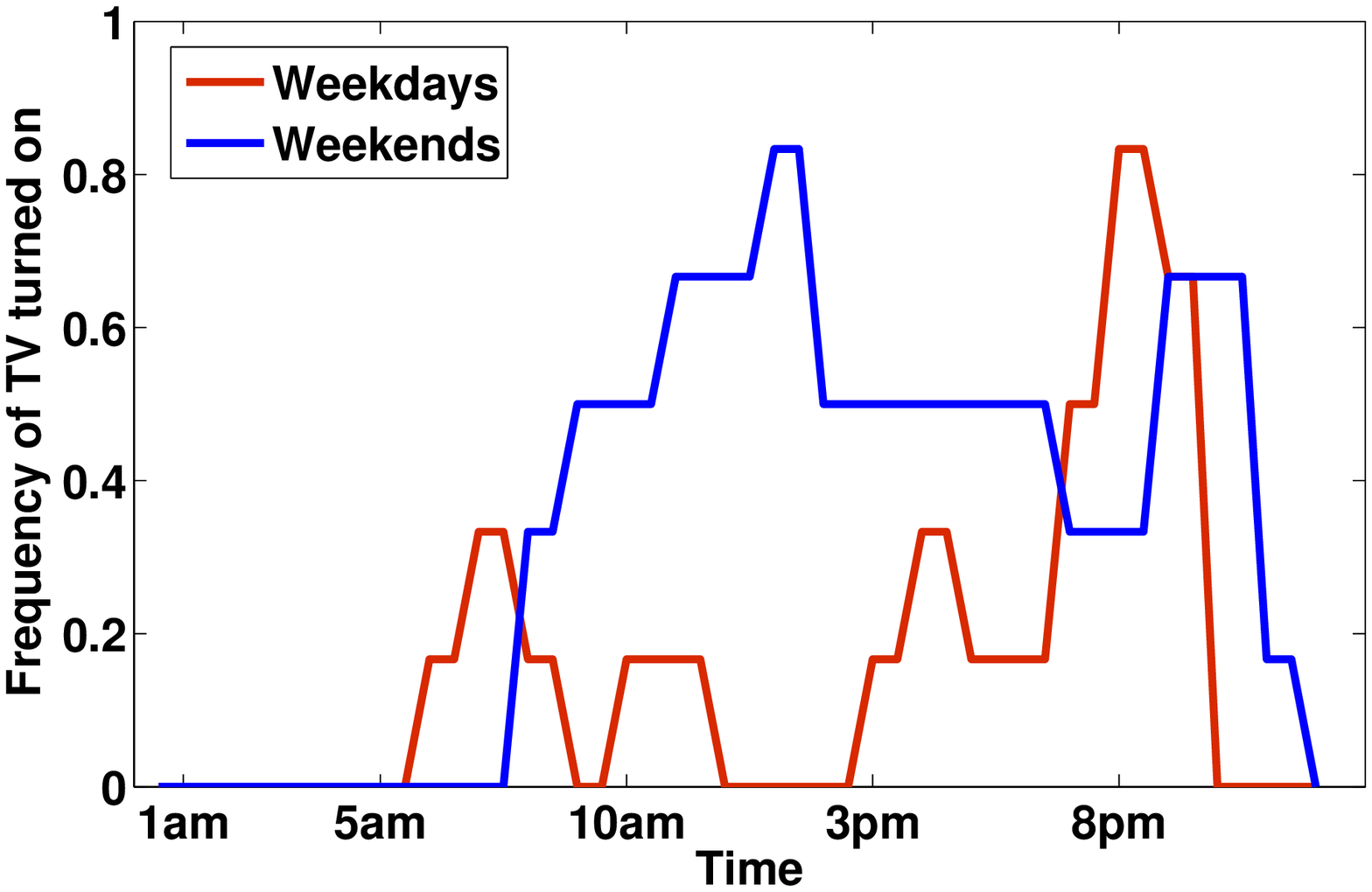}}\vspace{-0.2in}
\caption{Activity for a receiver using the broadcast through the air TV signals.} \label{fig:t-air}
}
\end{center}
\vspace{-0.3in}
\end{figure*}

{\bf Large Scale Activity Pattern Survey: }
The main results from the first survey are presented in Figures \ref{fig:t-weekdays} and \ref{fig:t-weekends}. We see that on weekdays people use their TV receivers mainly in the evening hours (e.g., after 6:30 p.m.). Late nights and morning times are fairly idle, hence, allowing opportunities for secondary access to spectrum.  However, this pattern changes on the weekends  to a more uniform distribution of the active periods, allowing for less secondary spectrum access opportunities. It is interesting to note the presence of many periods with low (if at all) activity. The period between 4:30 a.m. and 6:30 a.m. is completely inactive even during weekends.  
Delay tolerant applications could take advantage of these patterns and operate during the inactive periods, with high data rates. For instance, back up services on the cloud could run on scheduled times (e.g., every morning between 5 a.m. and 6 a.m.). Note here that even such applications would not be possible with the current spectrum sensing algorithms; primary signals would be sensed and no access would be granted to secondary users' applications.  

Some people might argue that the above levels of activity are still large. Hence, from their perspective, even with an {\em optimal} spectrum sensing algorithm that is able to identify active TV receivers, the secondary user would not have many chances to access the spectrum; the claim is that there are almost always, even small in numbers, some active receivers in a region.    
Thus, in our survey, we also asked the participants \textit{the type of service} they are using for their TV reception. Surprisingly, only 12\% of them actually use broadcast TV. The majority of the users - 77\% - use cable (and some use fiber and satellite). Since people that utilize cable and other technologies \textit{are unharmed by secondary spectrum access}, in Figure \ref{fig:t-air} we present the activity levels only for the users that actually receive the stations' signals over the air.  
The results clearly show that there are many more opportunities for secondary access to the spectrum (especially on weekdays). This might have been expected, knowing that this class of users is a very small portion of the total population. Note also that the broadcast users (even though we do not have demographics from the survey we conducted) might be located in areas where there is no cable or fiber service yet (or satellite is still expensive). These areas may correspond to isolated locations that might as well not be of interest for deploying a CR network (recall the earlier discussion about the value of a spectrum hole).

{\bf Apartment Block Case Study: }
The results from the Mechanical Turk survey reflect general patterns of users that may or may not be located in the proximity of each other. In order for a secondary link to be activated in a given location, {\bf all} the primary receivers in the vicinity should be not active. Even a single active primary device will force the secondary user to {\em backoff}. For instance, if all 20\% of the users that have their receivers turned on between 1 a.m. and 1:30 a.m. (see Figure \ref{fig:t-weekdays}) are spatially located close to each other (e.g., a neighborhood of a specific city, or an area of a specific state\footnote{For instance, there might be a TV show that is of extreme interest for these people and thus, it does not reflect the whole population in other areas of the country or even of the same city.}), then CR networks deployed in other areas will have spectrum opportunities during this time frame. On the contrary, if this 20\% is scattered across the country then the locations that will have these chances are fewer.  

\begin{figure}[ht]
\begin{center} 
\parbox{3in} { 
     \centerline{  
			\subfigure[Weekdays]{
				\includegraphics[scale=0.22]{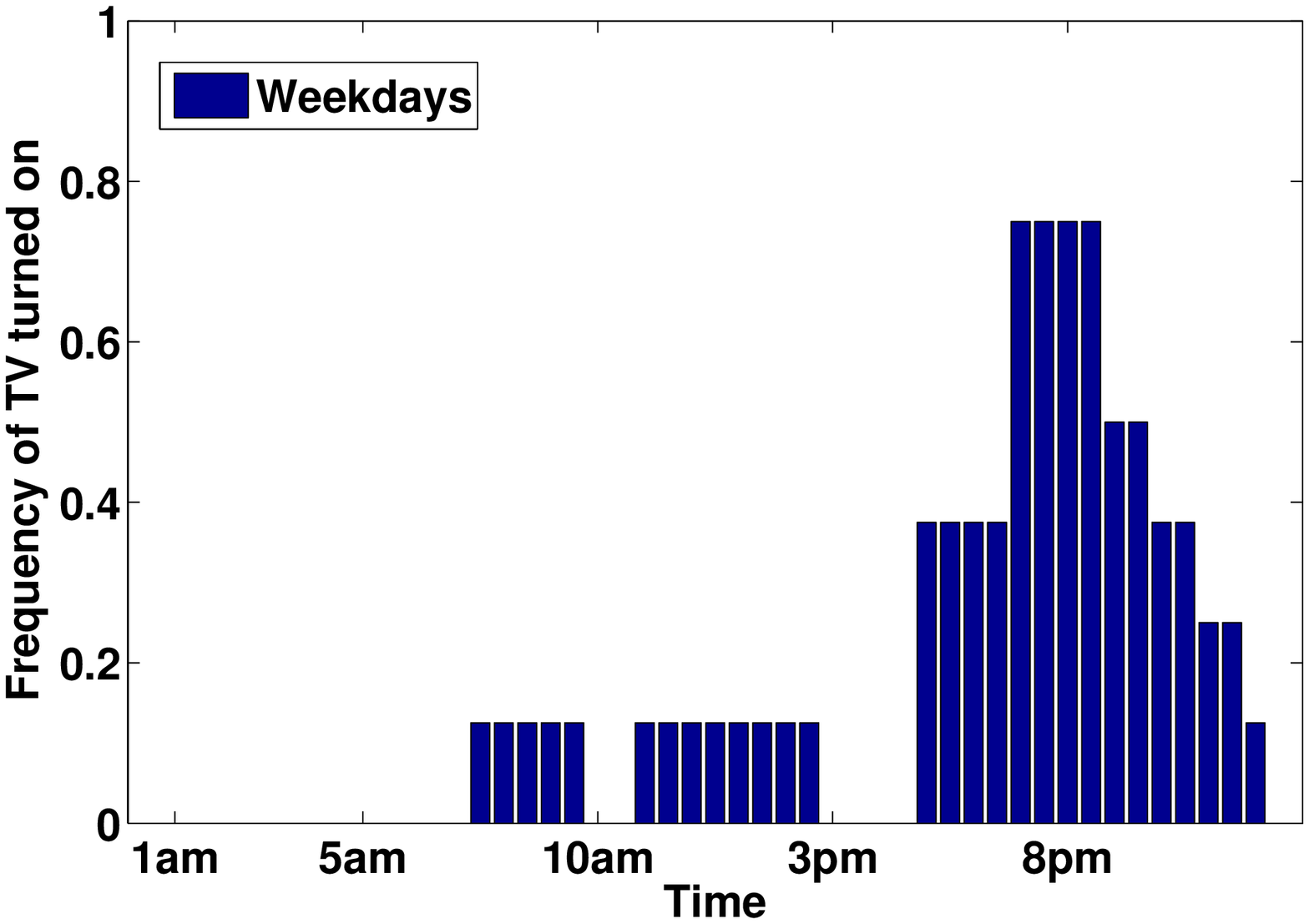} \label{fig:highland-weekdays}}\hspace{-0.2in}
			\subfigure[Reliable expert]{
				\includegraphics[scale=0.22]{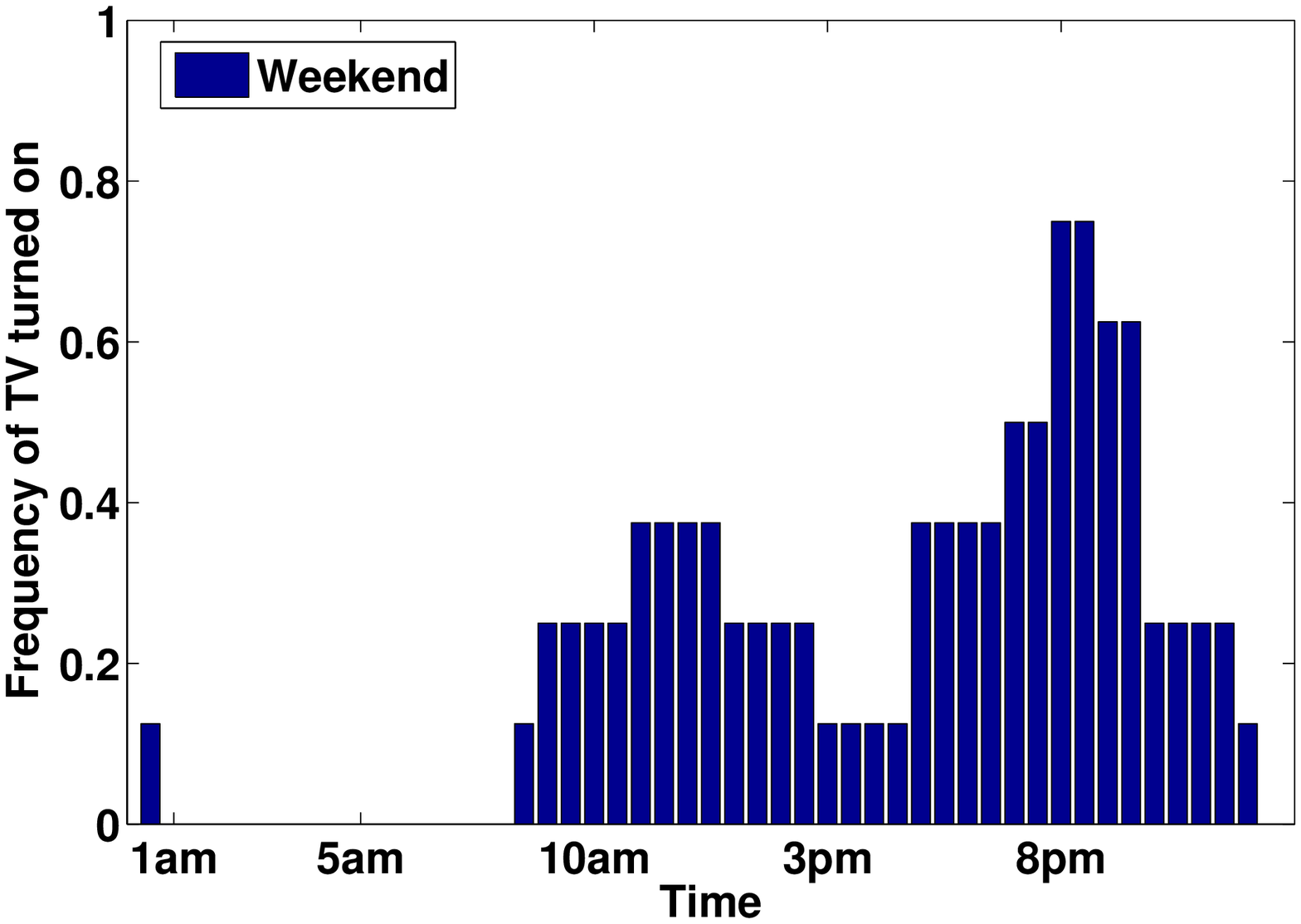}\label{fig:highland-weekends} }
	} \vspace{-0.2in}
    \caption{TV receiver activity in bins of 30 minutes for a specific block apartment building. }
    \label{fig:highland}
} 
\end{center}
\vspace{-0.3in}
\end{figure}

Thus the cumulative activity of primary co-located users is an important feature. To assess this quantitatively, we have performed the same survey with the residents of a block of apartments in a building consisting of 50 apartments. 
Figure \ref{fig:highland} depicts our results.


As we can see there are some clearly zero activity periods for the residents of this building, especially on weekdays. This cumulative activity of people allows many opportunities for a cognitive radio network to operate, even when current sensing algorithms would declare the presence of a primary transmitter. These opportunities are even larger, if we again consider only the activities of the people that \textit{actually} use the broadcast TV service, rather than cable, satellite, or fiber. Of the residents of the building under examination, only 4\% utilize  broadcast TV service and Table \ref{tab:highland} shows the union of the active periods for these users.

\begin{table}
\begin{center}
\begin{tabular}{|c|c|}
\hline Weekdays & Weekends \\ 
\hline 
\hline [7pm-9:30pm] & [9am-noon], [4pm-10pm] \\ 
\hline 
\end{tabular} 
\caption{TV receiver activity for the users of broadcast TV service of a specific block apartment building.}
\label{tab:highland}
\end{center}
\vspace{-0.3in}
\end{table}


{\bf Proposed Solutions: }
The passive nature of the receiver in many broadcast systems like television, can have a negative effect on the successful deployment of a CR network. Ideally the solution to the above problem would be to transform the passive receiver to an active one. In this way, we could sense the air for an {\em active reception} and decide to either access the medium or not. It is widely accepted in the research community that one of the properties that a proposed solution ``needs" to satisfy is to not involve large scale infrastructural changes. This is certainly a nice feature, but this is not the way that commercial operators always work. For example, very recently, a top US cable operator has required its customers to use free specific hardware in order to be able to continue receiving their service \cite{comcast} as it transitioned from analog to digital service. This is certainly something that no research paper would suggest.  
Nevertheless, this is the actual way that commercial operators work.  

Considering this, we ask why TV service providers should not equip every receiver with a ``box" that {\em informs} the CRs in its vicinity when the former is active? This could be perhaps paid for from the revenues from freeing the spectrum. Also note that only TV receivers that use the air need to be equipped (hence, even systems such as \cite{wild05} if not carefully deployed would not work efficiently since they do not distinguish between a cable TV or a broadcast receiver from the air). Further, the percentage of TVs that are equipped with some form of wireless interface (e.g., WiFi, Bluetooth etc.) has dramatically increased during the last years. We asked the participants in our surveys if their TV receiver is equipped with some type of wireless interface; 24\% of them answered ``Yes". These interfaces could also be utilized for similar purposes, reducing the cost for distributing the {\em special boxes} (just as some receivers are able to keep receiving digital TV signals without any modification). This would be a different type of sensing where specific messages in an unlicensed band would inform CRs about the existence of passive receivers.

A more drastic solution (which however requires many parties to negotiate and agree to corresponding service level agreements or SLAs -- e.g., FCC, providers etc.), is to completely abandon the broadcast TV service. Since only a small fraction of people utilize it, they could be served with other technologies for free (of course with the corresponding SLAs in place).  
This would free up the whole TV spectrum for usage by alternative wireless technologies.  
We recognize that both of the above solution sketches are radical and extreme, but in the authors opinion, only with drastic solutions will the vision of CR networks become viable.

\section{The FDD System Problem}
\label{sec:fdd_sys} 
\setcounter{paragraph}{0}

In this section we undertake a simplified analysis of the FDD uplink in a UMTS cell to demonstrate the feasibility of secondary transmissions even when the spectrum is presumably occupied. In CDMA systems \cite{Garg}, the energy per bit to interference ratio ($E_b/I$) at the base station receiver on a given 1.25 MHz (cdma2000) or 5 MHz (UMTS) channel needs to be between 5 and 7 dB and it is maintained nearly the same for all mobile stations communicating with the base station through strict power control. The value of $E_b/I$ on the uplink depends on the number of active phones $M$ in the cell, their voice activity $v_f$ (what fraction of time the user speaks), the power control accuracy $\eta_c$ and the fraction of interference from neighboring cells $f$. A simple expression for $E_b/I$ is $E_b/I = \frac{G_p \eta_c}{v_f (M-1)(1+f)}$ where $G_p$ is the processing gain (chip rate/bit rate) in CDMA through the use of spread spectrum. The maximum number of active mobiles on a channel $M_{max}$ can be roughly determined by setting $E_b/I$ in this expression to the minimum acceptable value. If we use some sample numbers such as $G_p = 3.84 Mcps/12.2 kbps = 315$, $v_f = 0.4$, $\eta_c = 0.8$, $f=0.75$, and the minimum $E_b/I = 3.2$ (5 dB), $M_{max} = 113$ per channel. As long as the $E_b/I$ value is maintained at the base station, and no mobile is denied service, there is no harm to the primary system. Service providers typically design their network to support peak capacity, which occurs infrequently, at specific hours/days of the week. At other times, the load is significantly smaller \cite{peng11} \cite{Tip10}. The operator can move users to other channels when a given channel reaches capacity.  

The question that arises is whether a secondary CR network can operate in the uplink frequencies  when the cellular network is not at capacity. By looking at the simple expression for $E_b/I$, this is indeed the case when $M < M_{max}$ \textit{\textbf{provided the interference caused by the secondary CR transmitter does not exceed the interference that may have been caused by the $M_{max} - M$ active mobile stations if they were present}}. A CR transmitter however cannot determine the corresponding transmit power by simply sensing the spectrum (even mobiles that are in idle mode occasionally transmit control data that may indicate that the spectrum is occupied although they have low duty cycles). A CR transmitter thus needs some information from the primary system (referred to as  collaboration in \cite{Peha10}) or has to use a worst case approximation. Also, it is not subject to power control and causes a constant interference at the base station if the CR transmitter is stationary. If the CR network has mobile transmitters and receivers, the situation becomes more complicated. Further the secondary receiver faces interference from the primary active mobile phones.


\noindent \textbf{Proposed Solutions:} The secondary network should use a transmission scheme that is based on spread spectrum and follows the 3G standards. It can operate by picking spreading codes fairly independently (since the uplink in CDMA systems use Gold or Kasami-like sequences to separate transmissions). With smartphones that can perhaps operate in an ad hoc mode and inexpensive hardware (similar to femtocells) that operate using 3G standards, a secondary network will be viable in this spectrum. 

\begin{figure}[htbp] 
   \centering
   \includegraphics[scale=0.18]{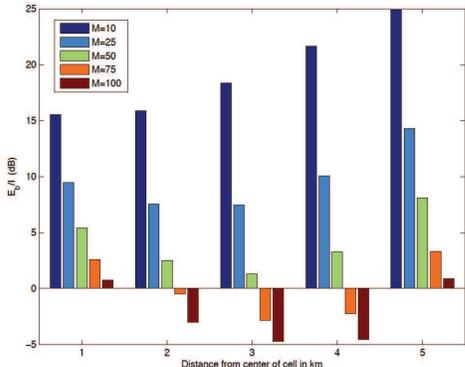} \vspace{-0.2in}
   \caption{$E_b/I$ at the secondary receiver for different numbers of active primary mobile stations in a cell.}
   \label{fig:sim}
   \vspace{-0.1in}
\end{figure}

The operating constraints on these networks is not very clear at this point. We can think of a CR network operating at close to the minimum transmit power to avoid causing interference at the base station irrespective of the location of the secondary user on the least used channels.  A study of mobile phone transmit powers in UMTS provides extensive data on the distribution of uplink transmit powers in a 3G network \cite{Raju08} . The data show that active mobile transmit powers can be as low as -50 dBm in certain scenarios (although they can be as high as 20 dBm). The average over several different scenarios (locations, network load, with and without Bluetooth) is 5.6 dBm (3.63 mW). 
This is however likely to restrict the throughput of the secondary network. If multiple secondary radio transmitters exist, the rates and transmit powers they use will be smaller. However cooperation with the primary network (which can share information about unused channels or channels with low loads) can allow the secondary network to operate more efficiently. For this, of course, either a market for secondary use or regulatory changes will become necessary.

To assess whether operating at a very small transmit power is viable for a secondary network, we perform a simple simulation. The transmit power of a primary mobile varies inversely as the path-loss because of strict power control. The signal strength from every mobile phone is thus received at approximately the same power S at the base station. The closer a mobile is to the BS, the lower is its transmit power. If the path loss exponent is $\alpha=4$ and the transmit power of a primary mobile at the cell edge (distance R from the base station) is $P_{edge}$, the transmit power at distance $r<R$ will be $\dfrac{P_{edge}r^4}{R^4}$.  Each of the $M$ active primary mobiles causes interference at the secondary receiver. Let the transmit power and distance of the $i$-th primary mobile from the secondary receiver be $P_i$ and $d_i$.  Then the overall interference including a voice activity of $u_f$ will be $u_f\displaystyle \sum_{i=1}^{M}\dfrac{P_i}{d_i^4}$. Assuming that the secondary transmitter is at a distance of 1 km from the secondary receiver and its transmit power is $P_s$, the $E_b/I$ observed at the secondary receiver will be\footnote{This simplified model ignores extra cell interference $f$.}:

\begin{equation}
E_b/I=\dfrac{G_pP_s}{u_f\displaystyle \sum_{i=1}^{M}\dfrac{P_i}{d_i^4}}
\end{equation}

We consider a cell of size $R=$ 5 km and place $M$ primary mobiles uniformly randomly in the cell.  We assume perfect power control with the transmit power of primary mobiles at 5 km from the base station to be 20 dBm. We place the secondary receiver at a distance of 1, 2, 3, 4, and 5 km from the base station. 
We let $P_s$ = 􀀀10 dBm, a fairly small number compared to the average transmit power of 5.6 dBm of primary mobiles in a cell
reported in \cite{Raju08}. We compute the average $E_b/I$ at the secondary receiver over 1000 runs for different values of $M$. The results are shown in Figure \ref{fig:sim}. We see that the secondary receiver has a reasonable $E_b/I$ close to the base station and at the edge of the cell, while it has poorer performance in between. This can be attributed to the fact that at the center of the cell, the interference comes from mobiles that are close to the base station and are transmitting at much lower powers. In fact, the secondary transmitter may have to reduce its transmit power at such locations beyond -10 dBm so as to not cause harm to the primary system. At the edge of the cell, the secondary receiver is far away from most of the active mobiles and sees lower interference. In the mid-regions, the secondary is not far away from mobiles that have a reasonably large transmit power and may not be able to operate unless $M$ is very small. 

However, this simple simulation demonstrates that there is spectrum access opportunity for secondary users in a cellular telephone network employing FDD under low loads, but the current approaches to dynamic spectrum access are not suitable for exploiting it.

\section{Scope of our Work \& Conclusions}
\label{sec:scope} 
\setcounter{paragraph}{0}

In this challenge paper we have tried to shed some light on the reasons behind the lack of commercial deployment of a cognitive radio network. In order to do so we focused on a specific functionality, important for the operations of such a network, that of spectrum sensing.  
However, we would like to emphasize on the fact that similar problems exist with respect to other functionalities, such as spectrum sharing and spectrum access. We emphasize that our work should not be viewed only as a question about the realization of cognitive radio networks. It should be seen more broadly as a challenge to the real world applicability of a large volume of existing research. For example, the research community is used to putting aside, in the majority of the cases, solutions that require large scale infrastructural changes. Without trying to argue in this paper whether this is correct or not, sometimes the only feasible solution(s) is(are) accompanied by this ``drawback". Thus, we should be more open to them and less critical of similar proposals. After all this is often the way that the world of commercial network operators functions.  

To sum up, solutions to technical problems might be much easier than we think. Coming back to the topic of this paper, CR networks, the nature of the primary user and the actual operations of a real communication system are usually neglected or made complicated or oversimplified. Therefore, the solutions proposed do not drive an effective commercially viable network architecture.  
Drastic solutions often need to be taken, but for this, a careful examination of the ``real" underlying tradeoffs need to be considered. For instance, in our sketch solution, for eliminating broadcast TV completely, the involved parties need to consider the potential benefits from the capacity/revenues obtained by freeing up the spectrum as well as the cost to serve people through other technologies (potentially for free).

\vspace*{1.5mm}
\bibliographystyle{unsrt}
\bibliography{main}


\end{document}